\documentclass[twocolumn,prc,aps,showpacs,amsmath,amssymb]{revtex4}
\usepackage{graphicx}

 \def\Pom{{ I\!\!P}}
 \def\Reg{{ I\!\!R}}
 \def\gsim{\mathrel{\rlap{\lower4pt\hbox{\hskip1pt$\sim$}}
 \raise1pt\hbox{$>$}}}

 \newcommand\la{\langle}
 \newcommand\ra{\rangle}
 \newcommand\beq{\begin{equation}}
 
 \newcommand\eeq{\end{equation}}
 \newcommand\beqn{\begin{eqnarray}}
 \newcommand\eeqn{\end{eqnarray}}
\def\mb{\,\mbox{mb}}

\def\GeV{\,\mbox{GeV}}

\def\lsim{\mathrel{\rlap{\lower4pt\hbox{\hskip1pt$\sim$}}
    \raise1pt\hbox{$<$}}}         
\def\gsim{\mathrel{\rlap{\lower4pt\hbox{\hskip1pt$\sim$}}
    \raise1pt\hbox{$>$}}}         

\def\mb{\,\mbox{mb}}

\def\GeV{\,\mbox{GeV}}

\def\s0{\sigma_0(s)}

\begin{document}

\title{\bf Perturbative fragmentation}

\author{B.Z.~Kopeliovich$^{1-3}$}
\author{H.-J.~Pirner$^{2,4}$}
\author{I.K.~Potashnikova$^1$}
\author{Ivan~Schmidt$^1$}
\author{A.V.~Tarasov$^{2,3}$}

\affiliation{$^1$Departamento de F\'\i sica
y Centro de Estudios
Subat\'omicos,\\ Universidad T\'ecnica
Federico Santa Mar\'\i a, Casilla 110-V, Valpara\'\i so, Chile\\
{$^2$Institut f\"ur Theoretische Physik der Universit\"at,
Philosophenweg 19, 69120
Heidelberg, Germany}\\
{$^3$Joint Institute for Nuclear Research, Dubna, Russia}
}

\date{\today}

\begin{abstract}

The Berger model of perturbative fragmentation of quarks to pions
\cite{berger} is improved by providing an absolute normalization and
keeping all terms in a $(1-z)$ expansion, which makes the
calculation valid at all values of fractional pion momentum $z$. We
also replace the nonrelativistic wave function of a loosely bound
pion by the more realistic procedure of projecting to the light-cone
pion wave function, which in turn is taken from well known models.
The full calculation does not confirm the $(1-z)^2$ behavior of the
fragmentation function (FF) predicted in \cite{berger} for $z>0.5$,
and only works at very large $z>0.95$, where it is in reasonable
agreement with phenomenological FFs. Otherwise, we observe quite a
different $z$-dependence which grossly underestimates data at
smaller $z$. The disagreement is reduced after the addition of pions
from decays of light vector mesons, but still remains considerable.
The process dependent higher twist terms are also calculated exactly
and found to be important at large $z$ and/or $p_T$.

 \end{abstract}


\maketitle

\section{Introduction}

The fragmentation of colored partons, quarks and gluons, into colorless
hadrons is an essential ingredient of any semi-inclusive hadronic
reaction, since confinement does not allow propagation of free color
charges. For this reason hadronization is usually considered to be related
necessarily to confinement specific to the string model \cite{cnn}.  
Indeed, the string model of hadron production is rather successful in
describing data.

In a typical event of quark fragmentation the mean production time
$t_p$ of a pre-hadron (i.e. a colorless cluster developing
afterwards a corresponding wave function) linearly rises with its
energy, and the most energetic hadron in such event takes about half
of the initial quark energy. In some rare events, however, the
leading hadron may take the main fraction $z\to1$ of the initial
quark energy. This process cannot last long, since the leading quark
is constantly losing momentum, $dp_q/dt=-\kappa$, where $\kappa$ is
the string tension. Therefore the production time should shrink at
$z\to1$ as \cite{kn},
 \beq
t_p=(1-z)\,\frac{E_q}{\kappa}\,.
\label{10}
 \eeq
 Notice that the end-point behavior of the production time, $t_p\propto
(1-z)$, is not specific for the string model, but is a result of
energy conservation.

\begin{figure}[htb]
 \includegraphics[width=6cm]{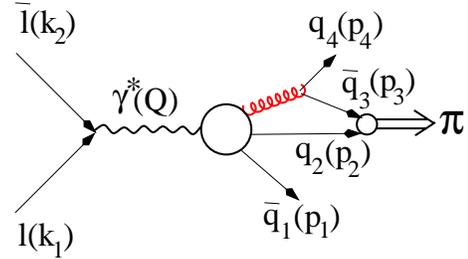}
\caption{\label{eebar} $l\bar l$ annihilation with production of two
$\bar qq$ pairs. The large blob contains gluon radiation by either
$\bar q_1$ or $q_2$. Four-momenta of particles are shown in parentheses.}
 \end{figure}

 The shortness of the production time is an indication that a
nonperturbative approach for the production of hadrons with large
$z\to1$ is not really required. Indeed, according to (\ref{10}), in this
region the hadronization time shrinks, i.e. the quark directly radiates
a hadron, $q\to h+q$. Furthermore, since the invariant mass squared of
the final state is $M_{qh}^2= m_h^2/z+m_q^2/(1-z)+p_T^2/z(1-z)$, where
$p_T$ is the transverse hadron momentum, at $z\to1$ the initial quark is
far off mass shell, and this process can be treated perturbatively. This
observation motivates a perturbative QCD calculations for leading pion
production $q\to\pi q$, within the model proposed by Berger
\cite{berger}, as is illustrated in Fig.~\ref{eebar} for $l\bar l$
annihilation.
 He found that the fragmentation function of a quark to a pion vanishes
as $(1-z)^2$ at $z\to1$, and falls as function of transverse pion
momentum as $1/p_T^4$. Besides, a nonfactorizable, scaling violating
term was found to dominate at $z\to1$. The shape of $z$-dependence
calculated by Berger \cite{berger} was found to agree well with data
after the inclusion of gluon radiation cf.  Ref.~\cite{kpps}.

Unfortunately, the calculation performed in \cite{berger} missed the
absolute normalization of the cross section, which makes it difficult
to compare with data. Moreover, it was done in lowest order in
$(1-z)$, therefore it is not clear in which interval of $z$ the
model is realistic. And last, but not least, the calculations were
based on the nonrelativistic approximation for the pion structure
function, assuming equal sharing of longitudinal and transverse
momenta by the quark and antiquark in the pion. However, the
dominant configuration of the $\bar qq$ pair projected to the pion
is asymmetric, with the projectile quark carrying the main fraction
of the momentum.

Here we perform calculations first in the Berger approximation, but
retaining the absolute normalizations and higher powers of $(1-z)$
(Sect.~3). Then, in Sect.~4 we give up the nonrelativistic
approximation and project the amplitude of $\bar qq$ production onto
the light-cone (LC) wave function of the pion. For this wave function
we consider three different models and find reasonable agreement
with phenomenological fragmentation functions (FF), but only at
large $z>0.95$. To improve agreement at smaller $z$ we add pions
originating from decays of $\rho$ and $\omega$ mesons, which are
produced by the same mechanism, and which is depicted in
Fig.~\ref{eebar}. In Sect.~5 we study higher twist contributions,
which gives a sizeable contribution in semi-inclusive pion
production in DIS at moderately large $Q^2$ and large $z$.

\section{Leading hadrons in Born approximation}

The amplitude of the process $l\bar l \to \bar q_1+q_2+G \to \bar q_1+q_2 + 
\bar
q_3+q_4$, depicted in Fig.~\ref{eebar}, in the lowest order of pQCD is given by,
 \beqn
&&A(l\bar l\to\bar q_1q_2\bar q_3q_4)
={1\over Q^2}\,
J_\mu^{(l)}(k_1,\bar\lambda_1;k_2,\bar\lambda_2)
\nonumber\\ &\times&
J_\mu^{(h)}(p_1,\lambda_1,i_1;p_2,\lambda_2,i_2;
p_3,\lambda_3,i_3;p_4,\lambda_4,i_4).
\label{100}
 \eeqn
 Here $k_1,\bar\lambda_1$ and $k_2,\bar\lambda_2$ are 4-momenta and
helicities of the lepton and antilepton respectively; $p_l,\
\lambda_l$ and $i_l$ are the 4-momenta, helicities and color indexes
of the quarks $q_1$ and $q_3$ ($l=1,3$) and antiquarks $\bar q_2$
and $\bar q_4$ ($l=2,4$).  The 4-momentum $Q=k_1+k_2$.

The leptonic and hadronic currents in (\ref{100}) read,
 \beq
J_\mu^{(l)}(k_1,\bar\lambda_1;k_2,\bar\lambda_2)=
e\,\bar u_{\bar\lambda_2}(k_2)\gamma_\mu
v_{\bar\lambda_1}(k_1)\,;
\label{120}
 \eeq
 \beqn
&&J_\mu^{(h)}(p_1,\lambda_1,i_1;p_2,\lambda_2,i_2;
p_3,\lambda_3,i_3;p_4,\lambda_4,i_4)
\nonumber\\ &=&
{1\over M^2}\,\sum\limits_{a=1}^{8}
{g_s\over2}\,\lambda^{(a)}_{i_2i_1}
{g_s\over2}\,\lambda^{(a)}_{i_4i_3}
\nonumber\\ &\times&
T_{\mu\nu}(p_1\lambda_1,p_2\lambda_2)\,
j_{\nu}(p_3\lambda_3,p_4\lambda_4)\,.
\label{140}
 \eeqn
 Here $M^2=(p_3+p_4)^2$ is the gluon invariant mass squared;
$g_s^2=4\pi\alpha_s$; $\lambda^{(a)}_{ij}$ are Gell-Mann matrices;
 \beqn
T_{\mu\nu}(p_1\lambda_1,p_2\lambda_2)&=&
e_{q_1}\,\bar u_{\lambda_2}(p_2)\,
\bigl[\gamma_{\mu}\hat G(Q-p_1)\gamma_{\nu}
\nonumber\\ &+&
\gamma_{\nu}\hat G(p_2-Q)\gamma_{\mu}\bigr]\,
v_{\lambda_1}(p_1)\,,
\label{160}
 \eeqn
 where $\hat G(q)=(\hat q+m_q)/(q^2-m_q^2)$; $\hat q=q_{\mu}\gamma_{\mu}$;
$m_q$ is the quark mass; and
 \beq
j_{\nu}(p_3\lambda_3,p_4\lambda_4)=
\bar u_{\lambda_4}(p_4)\,\gamma_\nu\,
v_{\lambda_3}(p_3)\,.
\label{170}
 \eeq

\section{Berger model}

In the Berger model \cite{berger} the amplitude $\tilde A$ of the
reaction $l\bar l\to\pi q_1q_4$ is a result of projection of the
amplitude Eq.~(\ref{100})  on the $S$-wave colorless state of the
$q_2\bar q_3$ pair having zero total spin. The result of the
projection is proportional to $\Psi_\pi(\vec r=0)$ ($\vec r$ is
3-dimensional) with a pre-factor $\sqrt{2/m_\pi}$ \cite{nemenov},
where $m_\pi$ is the pion mass. Then we get,
 \beq
\tilde A(l\bar l\to\pi q_1q_4)=
{1\over Q^2}\,J_\mu^{(l)}\,\bar J_\mu^{(h)}\,
\sqrt{2\over m_\pi}\,\Psi_\pi(0)\,.
\label{180}
 \eeq
 Here
 \beqn
&&\bar J_\mu^{(h)} = \frac{1}{\sqrt{3}}\sum\limits_{i=1}^3
\frac{1}{\sqrt{2}}\sum\limits_{\lambda=\pm1/2} {\rm sgn}(\lambda)
\nonumber\\ &\times& J_\mu^{(h)}(p_1,\lambda_1,i_1;p,\lambda,i;
p,-\lambda,i;p_4,\lambda_4,i_4)\,,
 \label{200}
 \eeqn
and the summations $\frac{1}{\sqrt{3}}\sum_{i=1}^3$ and
$\frac{1}{\sqrt{2}}\sum_{\lambda=\pm1/2}{\rm sgn}(\lambda)$ perform
projections to colorless and spinless states of the $q_2\bar q_3$
pair, respectively.

Then we can make use of the relations,
 \beqn
\sum\limits_{a=1}^8 \sum\limits_{i_1}^3
\lambda^{a}_{i_4 i}
\lambda^{a}_{i i_1} &=&
{16\over 3}\,\delta_{i_4i_1}\,;
\nonumber\\
\sum\limits_{\lambda=\pm1/2}
{\rm sgn}(\lambda)v_{-\lambda}(p_4)
\bar u_{\lambda}(p_3)\Bigr|_{p_3=p_4\equiv p}
 &=&
\gamma_5(\hat p+m),\nonumber\\
\label{220}
 \eeqn
 and arrive at the following form of the hadronic current,
 \beq
\bar J_{\mu}^{(h)} =
\frac{2g_s^2 e_{q_1}}{3\sqrt{6}\,M^2}\,
(j_{1\mu}+j_{2\mu})\,,
\label{240}
\eeq
 where
 \beqn
j_{1\mu}&=&
\bar u_{\lambda_4}(p_4)\,
\gamma_{\nu}\gamma_5(\hat p+m_q)
\gamma_{\nu}\hat G(Q-p_1)\gamma_{\mu}\,
v_{\lambda_1}(p_1)
 \nonumber\\&=&
\bar u_{\lambda_4}(p_4)\,\gamma_5
\left(\gamma_{\mu}-
\frac{2m_q\hat p\,\gamma_{\mu}}{M^2}\right)\,
v_{\lambda_1}(p_1)\,;
\label{260}
 \eeqn

 \beqn
j_{2\mu}&=&
\bar u_{\lambda_4}(p_4)\,
\gamma_{\nu}\gamma_5(\hat p+m_q)
\gamma_{\mu}\hat G(p_2-Q)\gamma_{\nu}\,
v_{\lambda_1}(p_1)
 \nonumber\\&=&
\frac{4}{Q^2-2pQ}\,
\bar u_{\lambda_4}(p_4)\,\gamma_5
\Bigl[(p_1p+m_q^2)\gamma_{\mu}
 \nonumber\\&-&
(p_{1\mu}+p_{\mu})\hat p -
m_q\hat p\,\gamma_{\mu} +
m_qQ_{\mu}\Bigr]\,
v_{\lambda_1}(p_1)\,.
\label{280}
 \eeqn
 Here we applied the algebra of $\gamma$-matrices, the Dirac equation and
4-momentum conservation, $Q=p_1+2p+p_4$. The invariant gluon mass $M$ was
defined in (\ref{140}).

It is convenient to choose the $z$-axis along the momentum $\vec
p_1$ in the collision c.m. frame, and to switch from Lorentz
4-vectors $a_{\mu}$ (e.g. $J_{\mu}^{(l,h)},\ p_{\mu1},\ p_{\mu4},\
Q_{\mu},\ etc.$) to light-cone vectors, $(a_+,a_-,\vec a_\perp)$,
where $a_{\pm}=a_0\pm a_z$. Since $\vec Q=0$, i.e. $Q_+=Q_-$, the
condition of gauge invariance, $Q_{\mu}J^{(l)}_{\mu}=Q_{\mu}\bar
J^{(h)}_{\mu}=0$, takes the form, $J^{(l)}_+=-J^{(l)}_-$ and $\bar
J^{(h)}_+=-\bar J^{(h)}_-$. Then the product of the lepton and
hadronic currents can be presented as,
 \beq
J^{(l)}_{\mu}\bar J^{(h)}_{\mu}=
-J^{(l)}_{+}\bar J^{(h)}_{+} -
J^{(l)}_{\perp}\bar J^{(h)}_{\perp}\,.
\label{300}
 \eeq

The typical values of transverse components are $|\vec p_\perp|\sim m_q$,
$\vec p_{4T}=-2\vec p$, $\left|\vec J^{(l)}_\perp\right|\sim J^{(l)}_+$, so
 \beq
\bar J^{(h)}_+\sim
{m_q\over Q}\left|\vec J^{(h)}_\perp\right|
\label{320}
 \eeq
Therefore, the first term in (\ref{300}) can be safely neglected.
Then we get,
 \beqn
\bar J^{(h)}_{\perp} &=&
\frac{4g_s^2e_{q_1}\delta{i_4i_1}}
{3\sqrt{6}\,M^2}\,
\bar u(p_4)\,\gamma_5
\nonumber\\ &\times&
\left[\xi\,\gamma_{\perp}-
\frac{2m_q}{M^2}\,\hat p\gamma_{\perp}\right]\,
v(p_1)\,,
\label{340}
 \eeqn
 where $\xi=(2+z)/(2-z)$, and
 \beq
z=\frac{p_{\pi+}}{Q_+}=\frac{2p_+}{Q_+}\,,
\label{380}
\eeq
 is the fractional pion momentum. In this approximation the invariant gluon
mass reads,
 \beq
M^2=(p+p_4)^2=
2\,\frac{m_q^2(1-z/2)^2+\vec p_{\perp}^2}
{z(1-z)}\,.
\label{360}
 \eeq

Notice that although the second term in (\ref{340}) is proportional
to the quark mass (which was assumed in \cite{berger} to be zero),
it should not be neglected. Indeed, after integration over $\vec
p_T$ the interference of the two terms in (\ref{340}) is of the same
order as the first term squared.

In this approximation the fragmentation function gets the form,
 \beqn
D_q^\pi(z)&=&\frac{64\alpha_s^2}{27m_\pi m_q^2}\,
\left|\Psi_\pi(0)\right|^2\,
\frac{z(1-z)^2}{(2-z)^2}
\nonumber\\ &\times&
\left[\xi^2+2(\xi+1)\left(\frac{z}{2-z}\right)^2 -
{16\over3}\,
\frac{z^2(1-z)}{(2-z)^4}\right].\nonumber\\
\label{400}
 \eeqn

The pion wave function at the origin correlates with the shape of the
parametrization for $\Psi_\pi(r)$. In the case of a Gaussian
parametrization,
 \beq
\left|\Psi_\pi(\vec r)\right|^2_{gauss}=
\frac{\kappa_1^3}{\pi^{3/2}}\exp(-\kappa_1^2 r^2/2)\,,
\label{420}
 \eeq
 the pion form factor has the form, $F_\pi(q^2)=\exp(-q^2/16\kappa_1^2)$. So
$\kappa_1^2=3/8\la r_{ch}^2\ra$.

With a bit more realistic exponential shape,
 \beq
\left|\Psi_\pi(\vec r)\right|^2_{exp}=
\frac{\kappa_2^3}{\pi^2}\exp(-2\kappa_2 r)\,,
\label{440}
 \eeq the pion form factor reads,
$F_\pi(q^2)=\left(1+q^2/16\kappa_2^2\right)^{-2}$. Then
$\kappa_2^2=2\kappa_1^2$.

These two examples demonstrate the high sensitivity of the wave function
at the origin to the choice of $r$-dependence. One finds
$|\Psi_\pi(0)|^2_{exp}/|\Psi_\pi(0)|^2_{gauss}=\sqrt{8\pi}\approx5$.
Therefore, it is difficult to conclude whether the Berger model agrees or not 
with data.

Another, more realistic option would rely on the pole form of the pion
form factor, $F_\pi(q^2)=\kappa_3^2/(\kappa_3^2+q^2)$, where
$\kappa_3^2=1/6\la r_{ch}^2\ra$. Then,
 \beq
\left|\Psi_\pi(\vec r)\right|^2=
\frac{1}{r}\exp(-\kappa_3 r)\,.
\label{460}
 \eeq
 In this case, however, the wave function at the origin is divergent.

 The Berger approximation, assuming that the pion production amplitude is
proportional to the amplitude of $\bar qq$ production with equal
momenta, would be justified if the pion was a nonrelativistic,
loosely bound system, i.e. $m_\pi\approx 2m_q$, $2m_q-m_\pi\ll m_q$.
However, the mean charge radius squared is much smaller than the
value given by such a nonrelativistic model, $\la
r_{ch}^2\ra=(4m_q^2-m_\pi^2)^{-1}$.

On the other hand, a description of the pion as a relativistic
bound system has been a challenge so far.

\section{Projection to the LC wave function}

\subsection{Direct pions}

In the light-cone (LC) representation the pion wave function depends
on the fractional LC momenta of the quark, $\alpha=p_{2+}/p_{\pi+}$,
and antiquark, $1-\alpha=p_{3+}/p_{\pi+}$, and the relative
transverse momentum, $k_\perp=\alpha
p_{3\perp}-(1-\alpha)p_{2\perp}$. In this representation the
amplitudes, Eqs.~(\ref{180}) and (\ref{100}), are related as,
 \beq
\tilde{A}=\frac{1}{(2\pi)^3}
\int\limits_0^1\frac{d\alpha}{\sqrt{2\alpha(1-\alpha)}}
\int d^2k_\perp\, A(\alpha,k_\perp)\,
\Psi_\pi(\alpha,k_\perp)\,,
\label{480}
 \eeq
 where the $\bar qq$ Fock component of the pion LC wave function
is normalized to unity,
 \beq
\int\limits_0^1 d\alpha\int d^2k_\perp\,
\left|\Psi_\pi(\alpha,k_\perp)\right|^2=1\,.
\label{500}
 \eeq

In this case the projection of the distribution amplitude of $q_2$ and
$\bar q_3$ on the pion LC wave function is more complicated that in
Berger model ($\alpha=1/2$), however it can be grossly simplified if one
neglects small terms of the order of $m$ and $k_\perp$ in comparison
with large $p_{2+}$ and $p_{3+}$ order terms. Then the combination in
Eq.~(\ref{220}) gets the simple form,
 \beq
\sum\limits_{\lambda=\pm1/2}
{\rm sgn}(\lambda)\,
v_{-\lambda}(p_3)\bar u_{\lambda}(p_2) =
\gamma_5\,\hat p_\pi+O(m,k_\perp)\,.
\label{520}
 \eeq
 Furthermore, neglecting small terms we arrive at a new relation for the
hadronic current of Eq.~(\ref{340}),
 \beqn
\bar J^{(h)}_{\perp} =
\frac{8g_s^2e_{q_1}\delta{\alpha_4\alpha_1}}
{3\sqrt{6}\,M^2}\,
\frac{1+(1-\alpha)z)}{1-\alpha z}\,
\bar u(p_4)\gamma_5\gamma_\perp
v(p_1).
\label{540}
 \eeqn
 When the momentum fractions of the quark and antiquark in the pion
wave function are $\alpha$ and $1-\alpha$, then the invariant mass 
squared reads,
 \beq
M^2=
\frac{m^2(1-\alpha z)^2+[(1-\alpha)\vec p_{\pi\perp}
-(1-z)\vec k_\perp]^2}
{z(1-z)(1-\alpha)}\,.
\label{560}
 \eeq

The light-cone pion wave function can be parametrized as
 \beq
\Psi_\pi(\alpha,\vec r) =
\phi(\alpha)\psi(r,\alpha)\,.
\label{740}
 \eeq
 If the wave function in momentum representation has a monopole form,
$\Psi_\pi(\alpha,k)\propto [k^2/\alpha(1-\alpha)+\kappa^2]^{-1}$,
then
 \beq
\psi(r,\alpha) =
N\,K_0(\kappa r\sqrt{\alpha(1-\alpha)}),
\label{750}
 \eeq
 where $K_0$ is the modified Bessel function.
Since the momentum dependence of $\Psi_\pi(\alpha,k)$ is poorly
known, we also performed calculations with a dipole dependent
wave function in the Appendix, since comparison of the results shows
the scale of the theoretical uncertainty.

The parameter $\kappa$ is fixed by the condition,
 \beq
-\frac{dF_\pi(q)}{dq^2}\Bigr|_{q^2=0}=
{1\over6}\,\la r_{ch}^2\ra\approx 1.83\GeV^{-2},
\label{760}
 \eeq
 where the pion form factor reads,
 \beq
F_\pi(q)=\int d^2r\int\limits_0^1 d\alpha
|\Psi_\pi(\alpha,\vec r)|^2\,
e^{i\alpha\vec q\cdot\vec r}\,.
\label{780}
\eeq

Thus, the parameter $\kappa$ as well as the normalization constant $N$ in
(\ref{740}) depend on the choice of function $\phi(\alpha)$.
We consider two popular models (compare with \cite{dijet}):

Model~1: Standard (asymptotic) shape \cite{asymp,radyushkin},
 \beqn
\phi_1(\alpha)&=&\alpha(1-\alpha)\,;
\label{790}\\
N_1^2&=&\frac{6\kappa_1^2}{\pi}\,;
\nonumber\\
\kappa_1^2&=&\frac{2}{\la r_{ch}^2\ra}\,.
\label{800}
 \eeqn

Model~2: Chernyak-Zhitnitsky model \cite{ch-zh},
 \beqn
\phi_2(\alpha)&=&\phi_1(\alpha)\,(1-2\alpha)^2\,;
\label{810}\\
N_2^2&=&\frac{70\kappa_2^2}{\pi}\,;
\nonumber\\
\kappa_2^2&=&\frac{6}{\la r_{ch}^2\ra}\,.
\label{820}
 \eeqn

To be specific we will calculate $D_{u}^{\pi^+}(p_T^2,z)$ which is
the FF of a $u$ quark into $\pi^+$. For the transverse momentum
dependent fragmentation function we have for each of these versions
(taking into account the longitudinal current contribution),
 \beqn
&&\left.\frac{dD_{u}^{\pi^+}(z,p_T^2)}{dp_T^2}
\right|_i =
2\left(\frac{\alpha_s}{2\pi}\right)^2
C_i\kappa_i^2z
\nonumber\\ &\times&
\left[(1-z)^2\,F_i^2(z,p_T)+
\epsilon z^2\frac{4p_T^2}{Q^2}\,
G_i^2(z,p_T)\right]\,.
\label{860}
 \eeqn
 Here $i=1,\ 2$; $C_1=1$; $C_2=35/3$;
 \beqn
F_i(z,p_T)&=&\int\limits_0^1 d\alpha\,
\frac{(1-\alpha)\phi_i(\alpha)}{\sqrt{a_i^2-b_i}}\
\frac{1+(1-\alpha)z}
{1-\alpha z}
\nonumber\\ &\times&
\ln\left(\frac{a_i+\sqrt{a_i^2-b_i}}
{a_i-\sqrt{a_i^2-b_i}}\right)\,;
\label{880}
 \eeqn

 \beqn
G_i(z,p_T)&=&\int\limits_0^1 d\alpha\,
\frac{(1-\alpha)^2\phi_i(\alpha)}
{(1-\alpha z)\sqrt{a_i^2-b_i}}\,
\nonumber\\ &\times&
\ln\left(\frac{a_i+\sqrt{a_i^2-b_i}}
{a_i-\sqrt{a_i^2-b_i}}\right)\,;
\label{900}
 \eeqn

 \beqn
a_i &=& p_T^2(1-\alpha)^2+m_q^2(1-\alpha z)^2
\label{920a}\\ &+&
\kappa_i^2\alpha(1-\alpha)(1-z)^2\,;
\nonumber\\
b_i &=& 4m_q^2\kappa_i^2(1-\alpha z)^2(1-z)^2
\alpha(1-\alpha)\,.
\label{920b}
 \eeqn

In fact, only the first leading twist term in square brackets in
(\ref{860}) corresponds to the factorized FF. The second term is a
higher twist term, whose value (factor $\epsilon$) is process
dependent, and which is discussed in more detail in Sect.~\ref{ht}
below .

\begin{figure}[htb]
 \includegraphics[width=7cm]{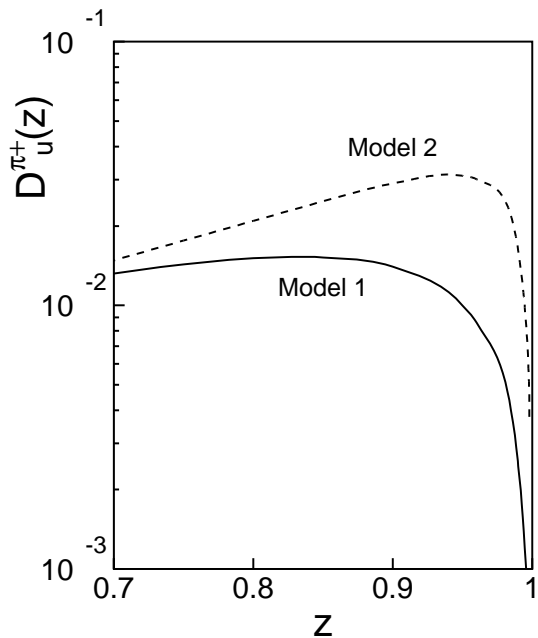}
 \caption{\label{3models} The fragmentation function,
 Eq.~(\ref{860}),
integrated over transverse momentum. Solid and dashed curves correspond to
the models $1$ and $2$ for the pion LC wave function (see text)
respectively. }
 \end{figure}

The results of the numerical calculations of the $p$-integrated FF,
for each of the three models, are plotted as functions of $z$ in
Fig.~\ref{3models}. The QCD coupling was fixed at $\alpha_s=0.4$.

The calculated fragmentation functions fall off with $z$ only at
very large $z\to1$, otherwise are rather flat, or even rise at small
values of $z$. Such a behavior does not comply with data which
suggest FF monotonically falling with $z$ \cite{{florian}}.
Apparently, the present calculations are missing some mechanisms
contributing at small $z$.

\subsection{Vector meson decays}

One of the processes contributing to the pion spectrum should be the
production, by the same mechanism shown in Fig.~\ref{eebar}, of
heavier mesons which decay to pions. One of the most important
corrections should come from $\rho$-meson production, which gives
the following contribution
 \beq
\Delta D_{u}^{\rho/\pi^+}(z)=
\frac{1}{\sqrt{1-\xi}}
\int\limits_{z_{min}}^1
\frac{dz'}{z'}\,\left[D_{u}^{\rho^+}(z')+
D_{u}^{\rho^0}(z')\right]\,.
\label{940}
 \eeq
 The bottom integration limit reads,
 \beqn
z_{min} &=& 2z\,\frac{1-\sqrt{1-\xi}}{\xi}\,;
\nonumber\\
\xi &=& \frac{4m_\pi^2}{m_\rho^2}\,.
\label{960}
 \eeqn

 We assume that $D_{u}^{\rho^+}(z)=3D_{u}^{\pi^+}(z)$, since $\rho$ has spin 1,
and that $D_{u}^{\rho^0}(z)={1\over2}D_{u}^{\rho^+}(z)$.

The $\omega$-meson production may also be important. Pions from
$\omega$ decays should be even softer because of the three-particle
phase space. The corresponding correction to the pion spectrum can
be calculated as follows.
 \beq
\Delta D_{u}^{\omega/\pi^+}(z)=
\frac{\int_{2m_\pi}^{m_\omega-m_\pi}
dM_{2\pi}\,g(M_{2\pi})\,I(z,M_{2\pi})}
{\int_{2m_\pi}^{m_\omega-m_\pi}
dM_{2\pi}\,g(M_{2\pi})}\,,
\label{962}
 \eeq
 where
 \beqn
g(M_{2\pi})&=&\sqrt{(M_{2\pi}^2-4m_\pi^2)
\left(\Omega^2-4m_\omega^2 m_\pi^2\right)}\,,
\nonumber\\
\Omega&=&m_\omega^2+m_\pi^2-M_{2\pi}^2\,;
\label{964}
 \eeqn
 and
 \beq
I(z,M_{2\pi})=\int\limits_{z_1}^{z_2}
\frac{dz'}{z'}\,D_{u}^{\omega}(z')\,,
\label{966}
 \eeq
 \beqn
z_1&=&{\rm min}\left\{1,\ \frac{2m_\omega^2\,z}
{\Omega +
\sqrt{\Omega^2-4m_\omega^2 m_\pi^2}}
\right\}\,,
\nonumber\\
z_2&=&{\rm min}\left\{1,\
z\,\frac{\Omega+
\sqrt{\Omega^2-4m_\omega^2 m_\pi^2}}
{2m_\pi^2}
\right\}\,.
\label{968}
 \eeqn
We assume that $D_{u}^{\omega}(z)= D_{u}^{\pi^+}(z)$, since the
factor of 3 coming from spin enhancement is compensated by an
isospin suppression.

\begin{figure}[htb]
 \includegraphics[width=7cm]{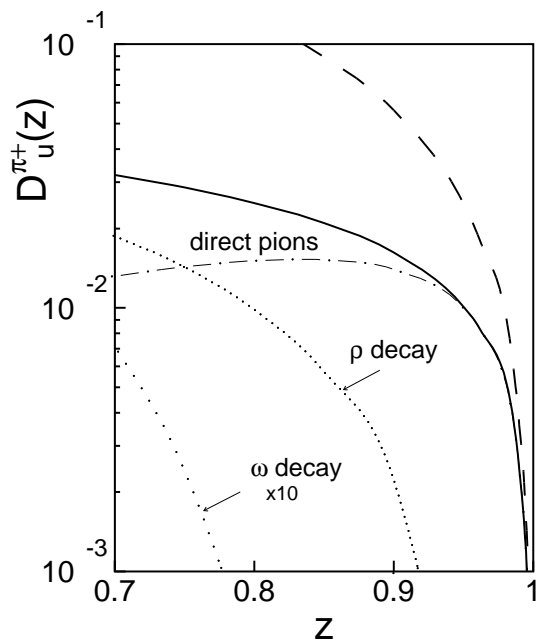}
 \caption{\label{pi-rho} Comparison of the Model~1 (asymptotic shape of
the pion wave function) with data. The curves from bottom to top are: {\it
dotted}: pions from $\omega$ and $\rho$ decays; {\it dot-dashed:} direct
fragmentation to pions; {\it solid:} sum of the three previous
contributions;  {\it dashed:} phenomenological FF for charged pions
\cite{florian} fitted to data at scale $\mu^2=0.5\GeV^2$.}
 \end{figure}

Fig.~\ref{pi-rho} shows our results for $D_{u}^{\pi^+}(z)$
(dashed-dotted), $\Delta D_{u}^{\rho/\pi^+}(z)$ and $\Delta
D_{u}^{\omega/\pi^+}(z)$ (dotted), and their sum (solid). We also
plotted the phenomenological $D_{u}^{\pi^+}(z)$ (dashed) obtained
from a global fit to data \cite{florian}.
 As anticipated, the production of $\rho$ contributes to the softer part of the
pion momentum distribution, and does not affect its hard part.

Other meson decays should pull the medium-$z$ part of $D_{u}^{\pi^+}(z)$
further up, but accurate calculation of all those contributions is still a
challenge.

Notice that our results have no $Q^2$ evolution, since the calculations are done
in Born approximation. Modification of the $z$-dependence by gluon radiation
makes it softer, closer to data, generating also a $Q^2$ evolution. These
corrections were studied within the Fock state representation in \cite{kpps}.

The transverse momentum distribution of pions is given by Eq.~(\ref{860}). One
cannot compare with data the mean value of $\la p_T^2\ra$ since it is poorly
defined. Indeed, $F_i\sim \ln(p_T)/p_T^2$ at high $p_T$ , so $\la p_T^2\ra$ is
divergent and depends on the upper cutoff.

\begin{figure}[htb]
 \includegraphics[width=7cm]{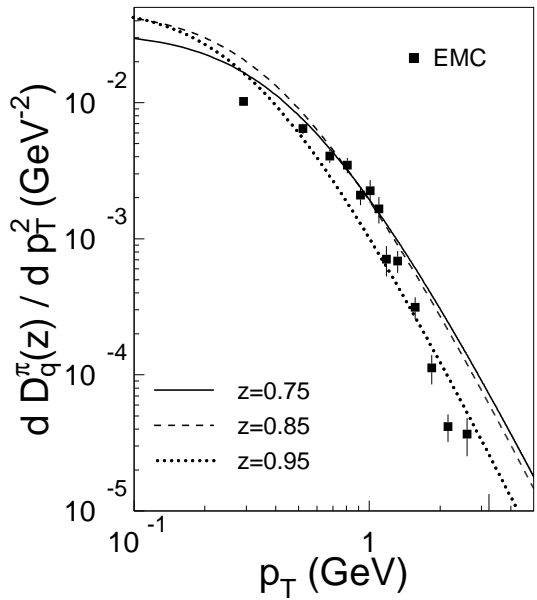}
 \caption{\label{pt} The transverse momentum dependent FF,
$dD_q^\pi(z)/dp_T^2$, calculated with Eq.~(\ref{860}) and the Model~1
for the production of direct pions. Solid, dashed and dotted curves are
calculated at $z=0.75,\ 0.85$ and $0.95$ respectively. Data from
\cite{emc} at $W^2>350\GeV^2$ are renormalized for a better
comparison with our results.}
 \end{figure}

Instead, one should compare with data the $p_T$ dependence. Our results for the
$p_T$-distribution of the FF, Eq.~(\ref{860}), is depicted in Fig.~\ref{pt} for
several values of $z$.
 
It might be too early to compare these results with data, since we did not
include yet the gluon radiation, intrinsic motion of quarks in the
target, and decays of heavier mesons. Nevertheless it is useful to check 
whether the calculated 
$p_T$
dependence is in a reasonable accord to data. Notice that the data
depicted in Fig.~\ref{pt} are integrated over a rather large $z$-bin,
$0.4<z<1$. The latter causes a considerable mismatch in normalization (see
Fig.~\ref{pi-rho}), so we renormalized the data \cite{emc} to be able to
compare the shapes, which then are in reasonable agreement.

\section{Higher twist terms}\label{ht}

The last term, in square brackets in Eq.~(\ref{860}), is a higher
twist effect. It does not vanish at $z\to1$, but is suppressed by
powers of $Q$. We neglected corrections of the order of $\la
p_T^2\ra/(zQ^2)$, which are important only at small $z$.

This higher twist term breaks down the universality of the
fragmentation function, since the factor $\epsilon$ depends on the
process. For $e^+e^-$ annihilation it is given by,
 \beq
\epsilon(l\bar l\to\pi\bar q_1q_4)=
\frac{sin^2\theta}{1+cos^2\theta}\,,
\label{980}
 \eeq
 where $\theta$ is the angle between the direction of $l\bar l$ collision and
momentum $\vec p_1$ in the c.m. frame.

For deep-inelastic scattering it reads,
 \beq
\epsilon(lq_1\to l'q_4\pi)=\frac{1-y}{2(1-y)+y^2}\,,
\label{1000}
 \eeq
 where $y=q_+/l_+$; $q_\mu$ is 4-momentum of the virtual photon;
$l$ is 4-momentum of the initial lepton.

The relative contribution of the higher twist term is,
 \beq
R_i(z,p_T)=4\epsilon\,
\left(\frac{z}{1-z}\right)^2
\frac{p_T^2\,G_i^2(z,p_T)}{Q^2\,F_i^2(z,p_T)}\,,
\label{1020}
 \eeq
 where subscript $i$ denotes the number of the model used for the LC pion wave
function, and $G_i$, $F_i$ are defined in (\ref{880})-(\ref{900}).

\begin{figure}[htb]
 \includegraphics[width=8cm]{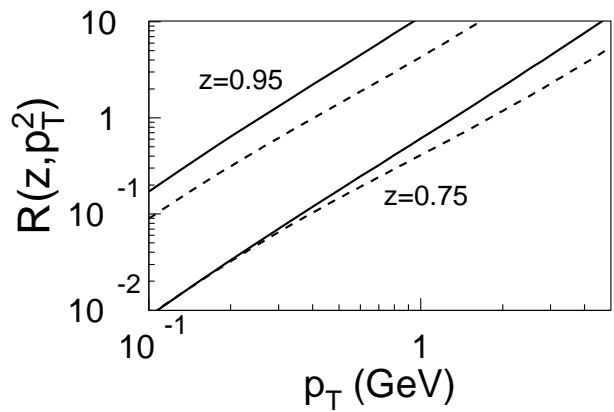}
 \caption{\label{h-twist-p} The relative higher twist correction to the FF
of a quark in DIS as function of transverse momentum for fixed values of
$z=0.75,\ 0.95$, and $Q^2=2.5\GeV^2$. The solid and dashed curves come
from calculations with Model~1, Eq.~(\ref{790}). and Model~2,
Eq.~(\ref{810}), respectively}
 \end{figure}

While the relative value of the nonfactorizable higher twist term is
expected to be vanishingly small in $l\bar l$ annihilation, it might
be a sizeable effect in SIDIS, usually associated with medium to
large values of $Q^2$. The relative correction, Eq.~(\ref{1020}), is
plotted in Fig.~\ref{h-twist-p} as function of $p_T$, for
$Q^2=2.5\GeV^2$ and several fixed values of $z$.
 Solid and dashed curve correspond to the models 1 and 2 for the LC pion
wave function, respectively. Although the higher twist term is relatively
small for forward fragmentation, it becomes a dominant effect at
$p_T^2\gsim 1\GeV^2$.

The corresponding higher twist correction to the $p_T$-integrated FF reads,
 \beq
R_i(z)= 4\epsilon\,\frac{\la p_T^2\ra}{Q^2}\,
\left(\frac{z}{1-z}\right)^2
\frac{\int_0^\infty dp_T^2\,G_i^2(z,p_T)}
{\int_0^\infty dp_T^2\,F_i^2(z,p_T)}\,,
\label{1040}
\eeq
 The factor $\la p_T^2\ra$ is divergent and depends on experimental
kinematic cuts. Therefore one should rely on its value specific for each 
experiment.

\begin{figure}[htb]
 \includegraphics[width=8cm]{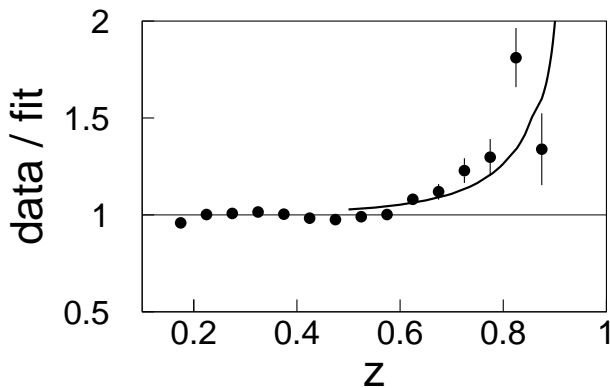}
 \caption{\label{ht-hermes} Hermes data \cite{hermes} for multiplicity of
charged pions produced in DIS on a proton, corrected for decays of vector
mesons. The data points are divided by the fit to the data at $z<0.65$
(see text). The curve corresponds to $R_1(z)+1$ calculated with
Eq.~(\ref{1040}) at $Q^2=2.5\GeV^2$ and $\la p_T^2\ra= 0.25\GeV^2$.}
 \end{figure}

Apparently, a direct way to see the higher twist contribution in data is
to study the $Q^2$ behavior of the FF. However, such data at sufficiently
large $z$ are not available so far. Therefore, we try to extract the
higher twist contribution from the $z$-dependence. To do so we first fit
data at moderate values $z<0.65$ where we do not expect a sizeable
higher-twist corrections, with the standard parametrization
$D_q^\pi(z)=Nz^\alpha(1-z)^\beta$. We use data from the HERMES experiment
\cite{hermes}.  We added the statistic and systematic errors in
quadratures.  The data are corrected by subtraction of the contribution
from diffractive vector mesons, $\gamma^*p\to\pi p$, which is another
higher twist contribution (see section \ref{3r-sect}).
We found $\alpha=-1.24\pm0.04,\ \beta=1.5\pm0.07,\ N=0.88\pm0.07$.
The data divided by this fitted $z-dependence$ are depicted in
In Fig.~\ref{ht-hermes} 
 We compare this data with the relative contribution of higher twists
$R_1(z)$, Eq.~(\ref{1040}), calculated at $Q^2=2.5\GeV^2$ and with the
measured value of $\la p_T^2\ra\approx 0.25\GeV^2$ \cite{hermes}. Our
results agree with the data reasonable well.

An attempt to see the higher twist effects in nuclear attenuation 
data was made in \cite{pg}. They found higher twist corrections of similar 
magnitude. 

Notice that other sources of pions, like decays of heavier mesons produced
via the same mechanism, are important for leading twist part. However,
they also supply the cross section with higher-twist terms. Nevertheless,
we assume that these corrections affect the ratio much less than the cross
section.

\section{Hints from triple-Regge phenomenology}\label{3r-sect}

\begin{figure}[htb]
 \includegraphics[width=8.5cm]{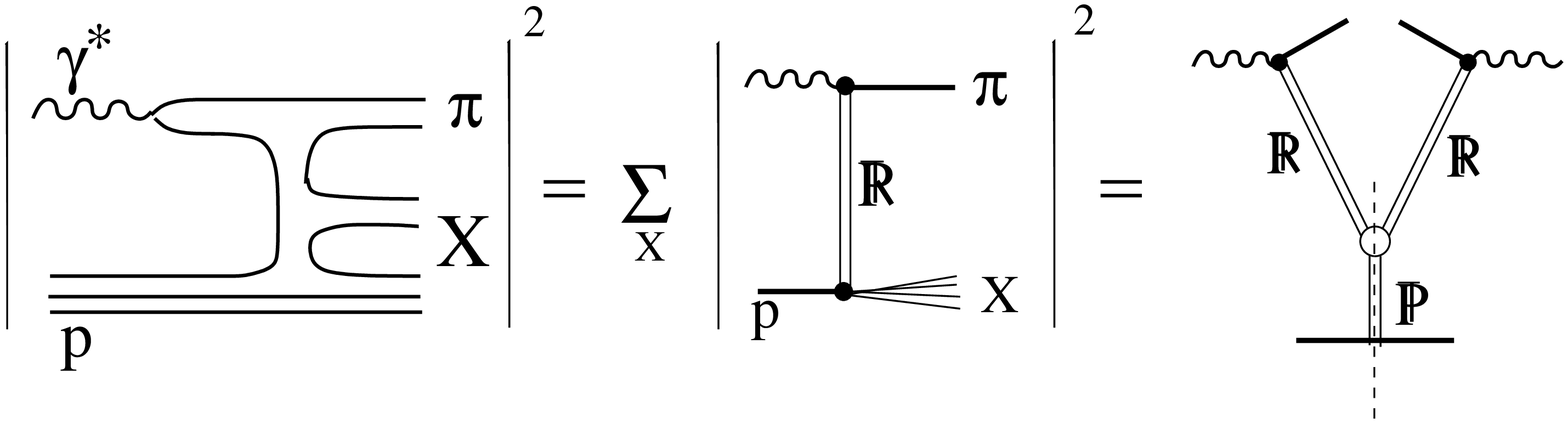}
\caption{\label{3r} Virtual photoproduction of a pion via Reggeon
exchange. The projectile quark from the photon fluctuation picks up
an antiquark, produced either from the vacuum or perturbatively (see
Fig.~\ref{eebar}), and they form a pion.}
 \end{figure}

The factorized part, Eq.~(\ref{400}), of the cross section of pion
production in $l\bar l$ annihilation, is the same as in
deep-inelastic scattering (DIS), where it can be compared with the
expectations of the triple-Regge description, illustrated in
Fig.~\ref{3r}.
 The inclusive cross section at fixed $z$ is energy independent (Feynman
scaling), and at fixed energy and $1-z\ll1$ depends on $z$ as,
 \beq
\frac{d\sigma(\gamma^*p\to\pi X)}{dz\,dp_T^2} \propto
(1-z)^n\,,
\label{462}
 \eeq
 where $z$ equals to Feynman $x_F$ in the triple-Regge kinematic region,
 \beq
z\approx x_F=\left(1-\frac{M_X^2}{s}\right)(1-x_{Bj})\,,
\label{463}
 \eeq
 and $x_{Bj}$ is the Bjorken variable.

The exponent in (\ref{462}) is related to the parameters of the
Regge trajectories involved,
 \beq
n=1-2\alpha_\Reg(p_T^2)\,.
\label{464}
 \eeq
 Here $\alpha_\Reg(p_T^2)$ is the trajectory of Reggeon $\Reg$.
The rapidity interval, $\Delta y \approx -\ln(1-z)$, covered by the
Reggeon is not large for the values of $z\sim 0.9$ under discussion.
Therefore the pion Regge pole should dominate, since it has large
coupling to nucleons. In this case, $\alpha_\pi(p_T^2)\approx
-\alpha^\prime_\pi p_T^2$, where $\alpha^\prime_\pi\approx
1\GeV^{-2}$. Thus,
 \beq
n_\pi = 2\alpha^\prime_\pi\,\la p_T^2\ra\approx 1.5\,.
\label{466}
 \eeq
 Here we rely on the value $\la p_T^2\ra\approx 0.25\GeV^2$ measured in
both HERMES \cite{hermes} and EMC \cite{emc} experiments. The value
of the exponent given in Eq.~(\ref{466}) agrees quite well with
data. Although our calculation confirmed the value $n=2$ found in
\cite{berger}, the inclusion of gluon radiation reduces the exponent
$n$ down to the value observed in data \cite{kpps}.

Notice that the $z$-dependence presented in
Eqs.~(\ref{462})-(\ref{464}) changes at very small $1-z\ll1$, and
becomes rather flat. Indeed, we assumed that the invariant mass
squared of the excitation $X$ is sufficiently large, $s(1-z)\gg
m_N^2$ for the Pomeron to dominate in the bottom leg of the triple
Regge graph in Fig.~\ref{3r}. However, this condition breaks down at
very small $1-z$ and Reggeons with $\alpha_\Reg(0)=1/2$ dominate in
the bottom leg. Another assumption we have made, pion dominance in
the $t$-channel exchange, is also violated when the rapidity
interval $\ln(1-z)$ becomes very large. Then Reggeons with a higher
intercept $\alpha_\Reg(0)=1/2$ become the dominant contribution.
Thus, the end-point behavior has the same power dependence,
Eq.~(\ref{462}), but with a different exponent,
 \beq
n(z\to1)=\alpha_\Reg(0)-2\alpha_\Reg(p_T^2)\approx
-{1\over2}+2\alpha^\prime_\Reg\,\la p_T^2\ra
\approx 0\,.
\label{468}
 \eeq

 Thus we arrive at the remarkable conclusion that the FF, which falls
steeply with $z$, levels off at very small $1-z\ll1$. This behavior,
dictated by the triple-Regge formalism, is more general than
perturbative calculations. One may wonder why this end-point feature
is absent in our calculations. What has been missed?  Notice that we
did not care about the fate of the recoil quark $q_4$ in
Fig.~\ref{eebar}, which was justified by the condition of
completeness. However, if the target excitation $X$ has a small
invariant mass, it affects the probabilities of different final
states of $q_4$.

The triple-Regge approach also indicates as an additional source of
a higher twist contribution, which is specific for semi-inclusive
DIS (SIDIS), the diffractive inclusive process $\gamma^*p\to\rho X$.
The $p_T$-integrated cross section corresponding to the
triple-Pomeron graph can be presented in the form,
 \beqn
\frac{d\sigma(\gamma^*p\to\rho X)}{dz} &=&
\frac{G^{pp}_{3\Pom}(0)/2\alpha_\Pom^\prime}
{(1-z)|\ln(1-z)|}\,
\frac{16\pi}{(\sigma_{tot}^{pp})^2}
\nonumber\\ &\times&
\left.\frac{d\sigma(\gamma^*p\to\rho p)}
{dp_T^2}\right|_{p_T=0}\,,
\label{1120}
 \eeqn
 where $G^{pp}_{3\Pom}(0)=3.2\mb/\GeV^2$ is the effective triple-Pomeron
coupling, extracted from the fit \cite{kklp} to data on $pp\to pX$.
Here we neglected the transverse size of the $\bar qq$ dipole
projected to $\rho$, since it is small, $1/Q^2$, and the $p_T$
dependence of the bare triple Pomeron vertex, since it is very weak
\cite{spots}. All the cross sections in (\ref{1120}) should be taken
at a c.m. energy squared $s^\prime=s_0/(1-z)$, where $s_0=1\GeV^2$.

The $z$-distribution of the produced $\rho^0$-mesons strongly peaks
at $z\to1$ (as any diffractive process should) and their decays feed
the effective FF $D_q^\pi(z)$,
 \beq
\left[\Delta D_{u}^{\rho/\pi^+}(z)\right]_{diff}=
\frac{1}{\sigma^{\gamma^*p}_{tot}}
\int\limits_{z_{min}}^1
\frac{dz'}{\sqrt{1-\xi}}\,
\frac{d\sigma(\gamma^*p\to\rho^0 X)}{z'dz'}\,.
\label{1140}
 \eeq
 Here $\xi$ and $z_{min}$ are defined in (\ref{960}). Due to color
transparency the amplitude of $rho$ production is inversely proportional
to $Q^2$, therefore $\sigma(\gamma^*p\to\rho^0 X)\propto 1/Q^4$. On the
other hand, the total virtual photoabsorption cross section is
$\sigma^{\gamma^*p}_{tot}\propto 1/Q^2$ (Bjorken scaling). Therefore, the
diffractive contribution to the effective FF $q\to\pi$ is a higher twist
effect, $\left[\Delta D_{u}^{\rho/\pi^+}(z)\right]_{diff}\propto 1/Q^2$.

The elastic production of vector mesons, $\gamma^* p\to Vp$
certainly also contributes to inclusive pion production, and is also
a higher twist effect. It can be evaluated using Eq.~(\ref{1140})
and a delta function for the $z'$-distribution of produced vector
mesons. However, in some cases, like in \cite{hermes}, this
contribution has been removed from data.

\section{Summary}

We performed calculations for the Berger perturbative mechanism
\cite{berger} of quark fragmentation into leading pions, keeping all
the sub-leading terms in powers of $(1-z)$ and all the coefficients.
Our results can be summarized as follows.

\begin{itemize}

 \item We performed a full calculation of the quark FF including higher
twist terms within the Berger approximation. However, we concluded
that the approximation of a nonrelativistic pion wave function is
unrealistic and brings too much uncertainty to the results of the
calculation.

\item
 We projected the produced $\bar qq$ pair distribution amplitude to the
light-cone pion wave function. For the latter we employed two
popular models: (i) the standard asymptotic shape (\ref{790}); (ii)
Model of Chernyak-Zhitnitsky (\ref{810}). Both models lead to a
$z$-dependence quite different from the one inferred from data. Only
at $z\geq 0.95$ our calculations agree reasonably with data (both
the shape and value), but greatly underestimate data at smaller
values of $z$.

\item
 Remarkably, the main amount of pions produced in quark fragmentation are
not produced directly, except the most energetic ones with $z>0.95$.
This fact should be taken into account in models employing
perturbative hadronization \cite{knph}

\item
 Searching for ways of improving the description of data we added
pions originated from decay of light vector mesons $\rho$ and
$\omega$. Although this contribution pulled up the production of
pions at medium to large $z$, apparently some contributions are
still missing. That may be production and decays of heavier mesons,
which are difficult to evaluate.

\item
 We also performed a full calculation for the higher twist term originated
from the longitudinal current contribution. It overcomes the leading twist
term at large $z$ and/or large transverse momenta.

\item
 A new higher twist contribution to pion production is found. It is
related to decays of diffractively produced vector mesons.

\end{itemize}

It worth reminding that our results for the FF at large $z>0.9$
should be compared with a phenomenological one with precaution.
First of all, data at such large $z$ are scarce and different
parametrizations \cite{florian,bkk,kkp} differ from each other
considerably. Second of all, our FF is calculated in the Born
approximation. Evolution (gluon radiation) may considerably change
the shape of the $z$-dependence \cite{kpps}.

\begin{acknowledgments}

We are grateful to Delia Hasch, Achim Hillenbrand and Pasquale Di Nezza
for providing us with the preliminary HERMES data. This work was supported
in part by Fondecyt (Chile) grants 1050519 and 1050589, and by DFG
(Germany)  grant PI182/3-1.

\end{acknowledgments}


 \def\appendix{\par
 \setcounter{section}{0}
\setcounter{subsection}{0}
 \def\thesection{Appendix \Alph{section}}
\def\thesubsection{\Alph{section}.\arabic{subsection}}
\def\theequation{\Alph{section}.\arabic{equation}}
\setcounter{equation}{0}}

 \appendix

\section{Dipole form of the pion LC wave function}
\label{dipole} \setcounter{equation}{0}

 \begin{figure}[h]
 \includegraphics[width=6cm]{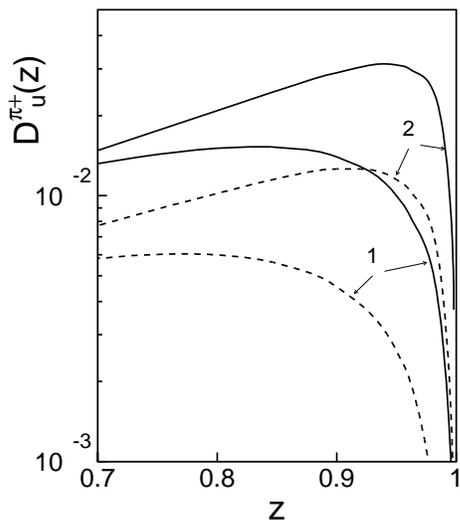}
 \caption{\label{pole-dipole} Fragmentation functions for direct pions
calculated with pole, Eq.~(\ref{740})  (solid curves), and dipole,
Eq.~(\ref{A740}) (dashed curves), parametrization for the transverse
momentum dependent part of the LC pion wave function. Labels $1$ and $2$
indicate the model used for the longitudinal momentum dependence of the
pion wave function.}
 \end{figure}

To see the sensitivity to the form $r$-dependence of the LC
wave function of the pion we also performed calculations with the dipole
parametrization of transverse momentum dependent part of the LC wave
function $\Psi_\pi(\alpha,\vec k)\propto
\left[\frac{k^2}{\alpha(1-\alpha)}+\kappa^2\right]^{-2}$. In impact parameter
representation it takes the form (compare with (\ref{740})),
 \beq
\Psi_\pi(\alpha,\vec r) =
N\,\phi(\alpha)\sqrt{\alpha(1-\alpha)}\,
rK_1(\kappa r\sqrt{\alpha(1-\alpha)}),
\label{A740}
 \eeq
 In this case we can still employ Eq.~(\ref{860}) for the fragmentation
function, but with a new form of function $F_i(z,p_T)$,
 \beqn
&&F_i(z,p)=\int\limits_0^1 d\alpha\,
\frac{(1-\alpha)\phi_i(\alpha)}{a_i^2-b_i}\
\frac{1+(1-\alpha)z}
{1-\alpha z}
\nonumber\\ &\times&
\left[a_i-2d_i+\frac{d_i(a-2e_i)}{\sqrt{a_i^2-b_i}}
\ln\left(\frac{a_i+\sqrt{a_i^2-b_i}}
{a_i-\sqrt{a_i^2-b_i}}\right)\right]\,,
\nonumber\\
\label{A880}
 \eeqn
 where $d_i=\kappa_i^2\alpha(1-\alpha)(1-z)^2$; $e_i=m_q^2(1-\alpha z)^2$.

Parameters $C_i$ and $\kappa_i$ in (\ref{860}) also get new values,

Model~1: asymptotic shape,
 \beqn
N_1^2&=&\frac{9\kappa_1^2}{2\pi};
\nonumber\\
\kappa_i^2&=&\frac{36}{5\la r_{ch}^2\ra};
\nonumber\\
C_1&=&3\,.
\label{A900}
 \eeqn

Model~2: Chernyak-Zhitnitsky shape,
 \beqn
N_1^2&=&\frac{105\kappa_2^2}{2\pi};
\nonumber\\
\kappa_i^2&=&\frac{108}{5\la r_{ch}^2\ra};
\nonumber\\
C_2&=&35\,.
\label{A920}
 \eeqn

The results of numerical calculations are depicted in Fig.~\ref{pole-dipole}
in comparison with calculations performed with the pole parametrization for the
pion wave function.


\end{document}